\documentclass[12pt]{article}

\begin{document}
\pagestyle{empty}
\begin{center}
{\Large \bf
The puzzle of the non-planar structure of the QCD contributions 
to the Gottfried sum rule}\footnote{Based on the invited  talks presented 
at the Workshop ``Hadron Structure in QCD: from Low to High energies'',
St.Petersburg, Repino, 18-22 May, 2004,  Russia and at the  Bogolyubov Conference 
``Problems of Theoretical and Mathematical Physics'', September 2-6, Dubna, 
Russia. Supported by RFBR Grants N 02-01-00601, 03-02-17047 and 
03-02-17177}

\vspace{0.1cm}

{\bf A.L. Kataev}\footnote{E-mail:kataev@ms2.inr.ac.ru }\\
\vspace{0.1cm}

Institute for Nuclear Research of the Russian  Academy of Sciences
,\\ 117312 Moscow, Russia  \\

\end{center}
\begin{center}
{\bf ABSTRACT}
\end{center}
 The recent finding, that in QCD at the $O(\alpha_s^2)$-level 
only non-planar diagrams, which are suppressed by a factor $1/N_c^2$ relative 
to planar ones, are contributing to the valence part of the Gottfried 
sum rule, is described. 
To our knowledge, this intriguing unique feature did not 
manifest itself previously in any perturbative expansion of any  gauge 
 quantum field model. We hope that this discovery  may have 
some theoretical explanation.
\noindent 
\vspace*{0.1cm}
\noindent
\vfill\eject

\setcounter{page}{1}
\pagestyle{plain}

\subsection*{1. Introduction.}
 The  studies of the lepton-nucleon deep-inelastic (DIS) sum rules 
provide important information about the structure of QCD both in perturbative 
and non-perturbative sectors. Moreover, their more detailed investigations 
are continuing to reveal new intriguing questions for further theoretical 
and phenomenological explanations. Among them are the results of the 
recent work of Ref. \cite{Broadhurst:2004jx}, which is devoted to the 
comparison of the obtained therein QCD predictions for the Gottfried 
sum rule \cite{Gottfried:1967kk} with the the expression for the Adler 
sum rule  \cite{Adler:1965ty} within the large-$N_c$ expansion.    

\subsection*{2.  
Gottfried and Adler sum rules: the definitions.}  
Consider first  the  isospin Adler sum rule, 
 expressed through the  structure function $F_2^{\nu N}$ of neutrino-nucleon 
 DIS as 
 \noindent
 \begin{equation}
 I_A=\int_0^1\frac{dx}{x}\bigg[F_2^{\nu p}(x,Q^2)-F_2^{\nu n}(x.Q^2)\bigg]
 =4I_3=2~~~.
 \label{eq:Adler}
 \end{equation}
In terms of parton distributions Eq. (\ref{eq:Adler}) takes the following form 
 \noindent
 \begin{equation}
 I_A=2\int_0^1dx[u_v(x)-d_v(x)]=2
 \label{eq:valence}
 \end{equation}
 where $u_v(x)=u(x)-\overline{u}(x)$ and $d_v(x)=d(x)-\overline{d}(x)$ are 
 the valence parton distributions of light quarks. 
It is possible to show, that 
$I_A$ 
receives neither perturbative nor non-perturbative $(1/Q^2)$-corrections 
\cite{Dokshitzer:1995qm}. In view of this the Adler sum rule  is 
$Q^2$ independent and demonstrates the consequence of  
the property  of scaling \cite{Bjorken:1968dy}.
This property (or  so called ``automodelling'' behaviour
of  structure functions) was rigorously proved  
 in the case of charged lepton-nucleon DIS  by N.N. Bogolyubov and coauthors  
 \cite{BVT} with application of general principles of local quantum field 
theory, described e.g. in the classical  text-book \cite{BSh}.
 However, it is known, that in QCD scaling is violated. 
The sources of its violation are related to the asymptotic freedom effects,
discovered within renormalisation-group concept \cite{RG} in the papers 
of the 2004 Nobel Prize laureates \cite{AF}, and to  non-perturbative 
contributions. Both types of these effects manifest themselves (though is some 
puzzling way) in the analog 
of the Adler sum rule, namely in the Gottfried sum rule. 
It can be  defined as the first $N=1$ non-singlet (NS)
Mellin moment of the difference of $F_2$ SFs of DIS of charged leptons on 
proton and neutron, namely   
\noindent
\begin{equation}
I_G^{v}=\int_0^1\frac{dx}{x}\bigg[F_2^{l p}(x,Q^2)-F_2^{l n}(x.Q^2)\bigg]
 =\frac{1}{3}\int_0^1 dx \bigg(u_v(x,Q^2)-d_v(x,Q^2)\bigg)~~.
 \label{eq:Got}
 \end{equation}
 The definition of Eq.~(\ref{eq:Got}) is presented in the case of 
assumption  accepted  previously  that the sea quarks are  
 flavour-independent.
 It corresponds to the condition $\overline{u}(x,Q^2)=\overline{d}(x,Q^2)$, 
 accepted in the early works on the subjects. 
However, due to the appearance  of  experimental data 
for the muon--nucleon DIS, Drell-Yan process  and semi-inclusive DIS 
 we know at present that this condition is violated and 
  $\overline{u}(x,Q^2)<\overline{d}(x,Q^2)$ ( for reviews see, 
 e.g. \cite{Kumano:1997cy}-\cite{ALK}). Therefore, the definition of 
 the  Gottfried sum rule should be modified as:  
 \noindent
 \begin{equation}
 I_G=\int_0^1\frac{dx}{x}\bigg[F_2^{l p}(x,Q^2)-F_2^{l n}(x.Q^2)\bigg]
 = I_G^v+\frac{2}{3}
 \int_0^1 dx \bigg(\overline{u}(x,Q^2)-\overline{d}(x,Q^2)\bigg)~~,
 \label{eq:Got1}
 \end{equation}
 where the last term has non-perturbative origin and is related 
to the manifestation of isospin-breaking effects in the Dirac sea.
 We will return to its discussion later on, after describing main 
puzzle, discovered in Ref. \cite{Broadhurst:2004jx}, 
 that in the $O(\alpha_s^2)$ level the scaling violation corrections to 
Eq.~(\ref{eq:Got}) have typically non-planar 
structure, namely in the large-$N_c$ limit 
 \cite{'tHooft:1973jz} (where $N_c$ is the number of colours)  
 they are suppressed by a $(1/N_c^2)$ factor. 
 This means, that the  leading in $N_c$ planar diagrams are cancelling 
 out in the analysed QCD corrections to the valence contribution 
 $I_G^v$. We hope that this discovery  may have 
some theoretical and phenomenological explanations.
 
\subsection*{3. Large $N_c$-expansion and 
the relation between Gottfried and Adler sum rules.} 
Let us now  support the statements made in the previous Section 
 by more formal considerations, presented in Ref. \cite{Broadhurst:2004jx}.
 The solution of the renormalization group equation for the valence 
 contribution $I_G^v$ to the Gottfried sum rule has the following 
 form
 \noindent
 \begin{equation}
 I_G^v= A(\alpha_s)C^{(l)}(\alpha_s)
 \label{eq:RG}
 \end{equation}
 with the anomalous-dimension term
 \noindent 
 \begin{equation}
 A(\alpha_s)=1+\frac{1}{8}\frac{\gamma_1^{(N=1)}}{\beta_0}\bigg(\frac
 {\alpha_s}{\pi}\bigg)+\frac{1}{64}\bigg(\frac{1}{2}\frac{(\gamma_1^{(N=1)})^2}
 {\beta_0}-\frac{\gamma_1^{(N=1)}\beta_1}{\beta_0^2}+\frac{\gamma_2^{(N=1)}}
 {\beta_0}\bigg)\bigg(\frac{\alpha_s}{\pi}\bigg)^2+O(\alpha_s^3)
 \label{AD}
 \end{equation}
 where $\beta_0$ and $\beta_1$ are 
 the first two scheme scheme-independent 
 coefficients of the QCD $\beta$-function, namely 
 \noindent
 \begin{eqnarray}
 \beta_0&=&\bigg(\frac{11}{3}C_A-\frac{2}{3}N_F\bigg) \\ 
 \label{beta_0}
 \beta_1&=&\bigg(\frac{34}{3}C_A^2-2C_FN_F-\frac{10}{3}C_AN_F\bigg)
 \label{beta1}
 \end{eqnarray}
 with $N_F$ active flavours and Casimir operators $C_F=(N_c^2-1)/(2N_c)$ and 
 $C_A=N_c$, in the fundamental and adjoint representation of $SU(N_c)$. 
The one-loop anomalous dimension term vanishes and the leading correction 
to  Eq.(\ref{AD}) comes from the scheme-independent two-loop contribution 
to the anomalous dimension function 
\noindent
\begin{equation}
\gamma_1^{N=1}=-4(C_F^2-C_AC_F/2)[13+8\zeta(3)-12\zeta(2)]
\label{gamma1}
\end{equation}
which was calculated in Refs.\cite{Ross:1978xk,Curci}. 
Notice the appearance of the distinctive non-planar colour factor 
$(C_F^2-C_AC_F/2)=O(N_c^0)$, which exhibits $O(1/N_c^2)$ suppression at 
large-$N_c$, in comparison with 
the individual weights of planar two-loop diagrams, 
namely  $C_F^2$ and $C_FC_A$, that are cancelling in the expression for 
$\gamma_1^{N=1}$. 

In the $\rm{\overline{MS}}$-like schemes the analytical expression 
for $\gamma_2^{N=1}$ was obtained in \cite{Broadhurst:2004jx} 
using a long-awaited determination of three-loop non-singlet splitting 
functions, made in 
Ref.\cite{Moch:2004pa}, and the results of the work \cite{r9}.
\newpage 

The result for  $\gamma_2^{N=1}$ reads \cite{Broadhurst:2004jx}:
\noindent
\begin{eqnarray}
\gamma_2^{N=1}&=&
(C_F^2-C_AC_F/2)\bigg\{
C_F\bigg[
290-248\zeta(2)
+656\zeta(3)
-1488\zeta(4)+832\zeta(5)
+192\zeta(2)\zeta(3)
\bigg]
\nonumber
\\&&{}
+C_A\bigg[
\frac{1081}{9}+\frac{980}{3}\zeta(2)-
\frac{12856}{9}\zeta(3)
+\frac{4232}{3}\zeta(4)
-448\zeta(5)
-192\zeta(2)\zeta(3)
\bigg]
\nonumber
\\&&{}
+N_F\bigg[
-\frac{304}{9}-\frac{176}{3}\zeta(2)+\frac{1792}{9}\zeta(3)
-\frac{272}{3}\zeta(4)\bigg]\bigg\} \\
\nonumber
&\approx&161.713785 - 2.429260\,N_F~~~~.
\end{eqnarray}
Notice the appearance in $\gamma_2^{N=1}$ of three non-planar factors,
namely $C_F^2(C_F-C_A/2)$, $C_FC_A(C_F-C_A/2)$ and $C_F(C_F-C_A/2)N_F$.
These results of Eq. (10) are generalising the observation 
of non-planarity of  $\gamma_1^{N=1}$-term of anomalous dimension function 
to three-loops and 
may be considered as the first non-obvious argument in favour 
of the correctness   of definite results of Ref. \cite{Moch:2004pa}. 

The additional perturbative contribution to Eq.~(\ref{eq:RG}) comes from 
radiative corrections to the coefficient function 
\noindent
\begin{equation}
C^{(l)}(\alpha_s)=\frac{1}{3}\bigg[1+C_1^{(l)N=1}
\bigg(\frac{\alpha_s}{\pi}\bigg)+ C_2^{(l)N=1}
\bigg(\frac{\alpha_s}{\pi}\bigg)^2+O(\alpha_s^3)\bigg]
\end{equation} 
where $C_1^{(l) N=1}$=0. The numerical expression for $C_2^{(l)N=1}$, namely 
\noindent 
\begin{equation}
C_2^{(l)N=1}=3.695C_F^2-1.847C_AC_F
\label{num}
\end{equation}
was obtained in Ref. \cite{Kataev:2003en} by numerical integration 
of the two-loop expression for the non-singlet coefficient function 
of the DGLAP equation \cite{DGLAP} calculated in the $x$-space  
in Ref. \cite{vanNeerven:1991nn}. Note, that 
it was not realized in Ref. \cite{Kataev:2003en} that 
Eq. (\ref{num}) has the  same non-planar structure as in the expression (9) 
for $\gamma_1^{N=1}$. This fact 
was demonstrated in Ref. \cite{Broadhurst:2004jx}, where the following 
analytical result for $C_2^{(l) N=1}$ was obtained: 
\noindent
\begin{equation}
C_2^{(l)N=1}=(C_F^2-C_AC_F/2)\bigg[-\frac{141}{32}+\frac{21}{4}\zeta(2)
-\frac{45}{4}\zeta(3)+12\zeta(4)\bigg]~~~~.
\label{anal}
\end{equation}
As was noticed by G. Grunberg, the $\overline{MS}$-scheme result 
for Eq. (\ref{anal}) is scheme-dependent. 

However, this observation 
does not affect the general feature of non-planarity of the $O(\alpha_s^2)$ 
correction to  $I_G^{v}$. Indeed, the transformation 
of the $\alpha_s$-corrections in  
Eq. (\ref{eq:RG}) to another 
$MS$-like scheme, which has the same expression 
for $\gamma_2^{N=1}$,  can be done with  the help of the shift 
\noindent
\begin{equation}
\frac{\alpha_s(Q^2)}{\pi}=\frac{\alpha^{'}_s(Q^2)}{\pi}+ \beta_0\Delta 
\bigg(\frac{\alpha_s^{'}(Q^2)}{\pi}\bigg)^2
\label{trans}
\end{equation}
where $\Delta$ is the concrete $N_c$-independent  number, 
which is defined by the 
logarithm  from the ratio of regularisation scales $\mu^2_{\overline{MS}}$
and $\mu^2_{MS-like}$. Thus, the general $\rm{MS}$-like scheme 
expression for the coefficient  $C_2^{(l) N=1}$
takes the following form
\noindent
\begin{equation}
C_{2~~ MS-like}^{(l)N=1}= C_{2~~\overline{MS}}^{(l)N=1}+\gamma_1^{N=1}\Delta
\label{shift}
\end{equation}
where both  $C_{2~~\overline{MS}}^{(l)N=1}$ and $\gamma_1^{N=1}$ have 
the same non-planar group weight $C_F(C_F-C_A/2)$. 
The transformation to other schemes, like MOM-schemes,  are more delicate. 
Indeed,  
they affect the value of $\gamma_2^{N=1}$ and may contain gauge-dependence.
In view of this we are avoiding their consideration. However, we hope, that 
these transformations will not spoil the non-planar structure 
of the $O(\alpha_s^2)$ approximation for $I_G^{v}$, found in 
Ref. \cite{Broadhurst:2004jx}.    
    
Taking into account the feature, that at $N_c\rightarrow\infty$ 
$\alpha_s/\pi=4/(\beta_0 ln(Q^2/\Lambda^2))$ and $\beta_0=(11/3)N_c$, 
we get the following expression for $I_G^{v}$ in the perturbative sector  
\noindent
\begin{equation}
I_G^{v}=\frac{1}{3}\bigg(1+O(1/N_c^2)\bigg)~~~~.
\label{IG}
\end{equation}
In the non-perturbative sector the ratio of the twist-4 $(1/Q^2)$-corrections 
of the Gottfried and say Gross-Llewellyn Smith sum rule \cite{Gross:1969jf}, 
defined as
\noindent
\begin{equation}
I_{GLS}=\int_0^1 \frac{dx}{x} \bigg[xF_3^{\nu p}(x,Q^2)+xF_3^{\overline
{\nu} p}(x,Q^2)
\bigg]~~~,
\label{GLS}
\end{equation}
can be estimated using renormalon calculus 
(for a review see e.g. \cite{Beneke:1998ui}). 
In the case of non-planar graphs, contributing to  $I_G$, 
the typical renormalon chain insertions into   the one-gluon line
of the quark-gluon ladder graph  should be crossed by the undressed 
second gluon line. As the result, 
it is expected in Ref. \cite{Broadhurst:2004jx} that 
at $N_c\rightarrow\infty$
the 
higher-twist contributions to the Gottfried sum rule 
is suppressed by a factor  
\noindent 
\begin{equation}
\frac{\alpha_s}{\pi N_c}\sim \frac{1}{N_c^2 ln(Q^2/\Lambda^2)}~~~.  
\label{ratio}
\end{equation}
relative to comparable effects in the Gross-Llewellyn Smith sum rule.

Using this estimate of Ref.\cite{Broadhurst:2004jx},  
one may  conclude, that in the limit of  $N_c\rightarrow\infty$
the Gottfried sum rule respects the isospin symmetry  both 
in perturbative and non-perturbative sectors and is related to the Adler 
sum rule as 
\noindent
\begin{equation}
I_G^{v}=\frac{2}{3}I_A\bigg(1+O(\frac{1}{N_c^2})\bigg)~~~~~.
\label{relation}
\end{equation}
However, in the real world, where $N_c=3$, there are experimental indications,
that in the nucleon sea there are isospin-breaking effects, which generate 
light-quark flavour asymmetry in the definite $x$-region and that 
$\overline{u}(x,Q^2)<\overline{d}(x,Q^2)$ (for a review of the 
developing experimental situation see Ref. \cite{Kumano:1997cy}-\cite{ALK}).
This, in its turn, necessitates the modification of the parton 
representation of the Gottfried sum rule following the definition 
of $I_G$ in Eq. (\ref{eq:Got1}). It is interesting, that the first indications 
to the violation of the quark-parton model prediction $I_G=1/3$ and the 
necessity of incorporation  of light-quark flavour asymmetry in  partonic 
language came from the results of rather old SLAC experiment of 
Ref.  \cite{Stein:1975yy}. However, the huge error-bars of these data 
and the appearance of EMC and BCDMS extractions of the Gottfried sum rule 
(see Refs. \cite{Aubert:1987da},\cite{Benvenuti:1989gs}), which 
gave no obvious 
indications to the existence of light-quark flavour asymmetry, resulted   
in the fact that in spite of the appearance of  
first theoretical considerations 
of the possibility that $\overline{u}(x,Q^2)\neq \overline{d}(x,Q^2)$
(see in particular the review of  Ref. \cite{Kumano:1997cy}),
this non-perturbative effect was not incorporated into early sets of parton 
distribution functions. At present this drawback is eliminated   
(see Refs. \cite{Gluck:1998xa}- \cite{Alekhin:2002fv}). Moreover, 
there are some additional theoretical arguments, which indicate 
that at $\overline{u}(x)<\overline{d}(x)$. Some of them, 
most related to the considerations of Ref. \cite{Broadhurst:2004jx},
will be discussed below.

\subsection*{4. Theoretical considerations}     
It is interesting that 
perturbative QCD considerations of Ref.\cite{Ross:1978xk}, which were based 
on the foundation, that the second coefficient 
of related anomalous dimension $\gamma_1^{N=1}$ is non-zero, were 
among first theoretical arguments of the 
existence of light-quark flavour asymmetry in the nucleon sea. 
However, 
noticed by the authors of \cite{Ross:1978xk} effect non-planarity of 
$\gamma_1^{N=1}$ was not related to large-$N_c$ expansion language. 
Unfortunately, this important work of  Ref.\cite{Ross:1978xk}, which 
contributed to the understanding of the necessity of introduction of 
light-quark flavour asymmetry in the parton distributions, was also 
forgotten in the definite moment (probably, the effect discussed 
in Ref.\cite{Ross:1978xk} was considered to be numerically not 
essential). 
 
Other theoretical evaluations   of   
light-quark flavour asymmetry contributions 
were discussed in the review reports 
of Ref. \cite{Kumano:1997cy},\cite{GP}. Here  we will mention
the works, where the   non-perturbative  QCD methods were 
used. 
Among these methods is the developed in  Ref. \cite{Dorokhov:1993fc}  
instanton 
model  and essentially based on the 
large-$N_c$ expansion chiral soliton model of Ref. \cite{Diakonov:1996sr}, 
which was used  in Ref. \cite{Pobylitsa:1998tk}. 
in estimates of the measure of light-quark flavour asymmetry in the nucleon 
sea.  

Since in Ref. \cite{Broadhurst:2004jx} and in  some other 
discussions presented above
large-$N_c$ expansion approach was essentially used,   
it is reasonably to think that the considerations  
of Ref. \cite{Broadhurst:2004jx}
and Ref. \cite{Pobylitsa:1998tk} may be compatible.
In the letter case the values of $I_G$ between 0.219 and 0.178 were obtained 
for a range of constituents quark mass between 350 and 420 ${\rm MeV}$,
in fair agreement with the announced NMC result $I_G^{exp}=0.235 \pm 
0.026$ at $Q^2=4$ GeV$^2$ \cite{Arneodo:1994sh}. These values for $I_G$ 
are essentially
based on 
contribution 
\noindent
\begin{equation}
\frac{1}{2}(3I_G-1)=\int_0^1 dx \bigg(\overline{u}(x)-\overline{d}(x)\bigg)
=O(N_c^0)~~~~.
\label{relation1}
\end{equation}
 estimated in Ref. \cite{Pobylitsa:1998tk}.
It is worth to note, that for the constituent quark mass $M$=350 MeV 
the
$x$-behaviour for the difference of $x[\overline{u}(x)-\overline{d}(x)]$ 
turned out to be in rather good agreement with the $x$-behaviours of this 
quantity calculated with the help of  next-to-leading order (NLO)  
GRV parameterisation. 
Thus, at the NLO level one is able to describe at the qualitative level 
the existence of light-quark flavour asymmetry using the method   
of Ref. \cite{Pobylitsa:1998tk}, essentially based on the large-$N_c$ 
expansion.

The comparison of the results of Ref. \cite{Broadhurst:2004jx} 
with the ones of Ref. \cite{Pobylitsa:1998tk} generate several 
interesting to our mind questions. In conclusion let us  mention 
several ones.
\begin{enumerate}
\item Does typical non-planar structure of the perturbative 
series for $I_G^{v}$, observed at the $O(\alpha_s^2)$-level,
is continuing to manifest itself in higher orders ?
\item What are theoretical and possible phenomenological consequences 
of the regular non-planar structure of the considered perturbative series ?
\item Is there any theoretical relation between large-$N_c^2$ suppressed 
results of Ref. \cite{Broadhurst:2004jx} and the existence of light-quark 
flavour asymmetry in the nucleon sea ?
\item What is the real value of the measure of light-quark flavour asymmetry, 
defined in Eq. (\ref{relation1}) ?
\item 
Does this non-perturbative quantity is $Q^2$-dependent ?
\end{enumerate}
Future will show, whether it will be possible to find answers on at least some 
of the questions given above and thus to understand the perturbative QCD 
puzzle, discovered in Ref. \cite{Broadhurst:2004jx}.

I would like to thank D. Broadhurst and and C.J. Maxwell for the  
enjoyable  collaboration, which resulted in the work 
of Ref.\cite{Broadhurst:2004jx}, and for further useful 
discussions and exchange of information. It is a pleasure to thank 
J. Blumlein, G. Grunberg,  A.B. Kaidalov, W. van Neerven,  G. Parente and 
C. Weiss 
for helpful conversations, related to the subjects, described in this report.
This   work   was supported by the 
RFBR Grants N 02-01-00601, 03-02-17047 and 03-02-17177.
The material, related to this contribution, was presented 
in the invited  talks at 
the Workshop ``Hadron 
Structure and QCD:from Low to High energies'', St. Petersburg, Repino,
Russia,18-22 May, Russia  and at  the Bogolyubov Conference ``Problems 
of Theoretical and Mathematical Physics'', September 2-6, Dubna, Russia, 
I would like to thank Organizers of these important events
for creating productive scientific atmosphere 
and for the financial support.

\end{document}